# Generalized Polarization and time-resolved fluorescence provide evidence for different populations of Laurdan in lipid vesicles

Mihaela Bacalum[1], Mihai Radu[1], Silvio Osella[2], Stefan Knippenberg[3,4] and Marcel Ameloot[3,*]

[1] Department of Life and Environmental Physics, Horia Hulubei National Institute for Physics and Nuclear Engineering, Reactorului, 30, Măgurele, 077125, Romania.

[2] Chemical and Biological Systems Simulation Lab, Centre of New Technologies, University of Warsaw, Banacha 2C, 02-097 Warsaw, Poland.

[3.] Biomedical Research Institute, Hasselt University, Agoralaan Bldg. C, 3590 Diepenbeek, Belgium.

[4] Theory Lab, Hasselt University, Agoralaan Bldg. D, 3590 Diepenbeek, Belgium

Corresponding author:
Dr. Marcel Ameloot
Hasselt University, BIOMED,
Agoralaan, gebouw C, Diepenbeek, 3590, Belgium,
tel, +32-11-269233; fax, +32-11-269299,
e-mail: marcel.ameloot@uhasselt.be

**Abstract**. The solvatochromic dye Laurdan is widely used in sensing the lipid packing of both model and biological membranes. The fluorescence emission maximum shifts from about 440 nm (blue channel) in condensed membranes ($S_o$) to about 490 nm (green channel) in the liquid-crystalline phase ($L_\alpha$). Although the fluorescence intensity based generalized polarization (*GP*) is widely used to characterize lipid membranes, the fluorescence lifetime of Laurdan, in the blue and the green channel, is less used for that purpose. Here we explore the correlation between *GP* and fluorescence lifetimes by spectroscopic measurements on the $S_o$ and $L_\alpha$ phases of large unilamellar vesicles of DMPC and DPPC. A positive correlation between *GP* and the lifetimes is observed in each of the optical channels for the two lipid phases. Microfluorimetric determinations on giant unilamellar vesicles of DPPC and DOPC at room temperature are performed under linearly polarized two-photon excitation to disentangle possible subpopulations of Laurdan at a scale below the optical resolution. Fluorescence intensities, *GP* and fluorescence lifetimes depend on the angle between the orientation of the linear polarization of the excitation light and the local normal to the membrane of the optical cross-section. This angular variation depends on the lipid phase and the emission channel. *GP* and fluorescence intensities in the blue and green channel in $S_o$ and in the blue channel in $L_\alpha$ exhibit a minimum near 90°. Surprisingly, the intensity in the green channel in $L_\alpha$ reaches a maximum near 90°. The fluorescence lifetimes in the two optical channels also reach a pronounced minimum near 90° in $S_o$ and $L_\alpha$, apart from the lifetime in the blue channel in $L_\alpha$ where the lifetime is short with minimal angular variation. To our knowledge, these experimental observations are the first to demonstrate the existence of a bent conformation of Laurdan in lipid membranes, as previously suggested by molecular dynamics calculations.

**Introduction**

Laurdan (6-dodecanoyl-2-dimethylamino naphthalene) is one of the most extensively used solvatochromic dyes to explore both model and natural membranes properties due to a good partitioning from water to lipid membranes, even partition to different lipid phases and good sensitivity to lipid packing. [1-7] When inserted into lipid membranes, the fluorescence properties of Laurdan depend on the local polarity and the relaxation of the water molecules near the excited naphthalene moiety, which is generally considered to be located at the level of sn-1 carbonyl of the glycerol backbone of the phospholipids.[1, 8] The emission maximum of Laurdan shifts from around 440 nm in the gel phase ($S_o$) to around 490 nm in the liquid-crystalline phase ($L_\alpha$). [1, 9] These shifts from the blue to the green emission band are ascribed to the relaxation of the water molecules around the increased dipole moment of the probe in the excited state,[9, 10] so that the emission in the blue channel is associated with the nonrelaxed form of Laurdan and the emission in the green channel with the relaxed form.[2, 11, 12]

The spectral shifts can be quantified by the so-called generalized polarization (*GP*),

$$GP = \frac{I_{440} - I_{490}}{I_{440} + I_{490}} \qquad (1)$$

where $I_{440}$ and $I_{490}$ are the emission intensities at 440 and 490 nm, respectively.[3] *GP* has been proven to be an adequate tool for characterizing lipid phases in small and large unilamellar vesicles,[3, 13-15] giant unilamellar vesicles (GUVs) and cell membranes.[5, 16-20] Leung *et al.* demonstrated a correlation between *GP* and the $^2$H NMR order parameter for various lipid membranes in the liquid crystalline state.[21]

Fluorescence microscopy allows the mapping of *GP* over the lipid membrane by assigning a *GP* value to each pixel. This can be implemented by using two optical channels at the detection side or a spectral array detector. In the latter case the so-called spectral phasor approach can be utilized to quantify both changes in wavelength and spectral width for each pixel.[2, 17]

A rigorous approach in the study of the solvent relaxation process in the lipid membrane requires monitoring of the time-dependent fluorescent shifts on the nanosecond time scale.[22] Because of practical reasons, many studies determine the fluorescence relaxation time in the whole emission band or in the blue and the green emission band only.[4, 6, 23-25] Although strictly speaking the observed relaxation times are not fluorescence lifetimes because of the solvent relaxation effects, these are denoted so in most reports, as will be done here as well. Lifetime spectroscopy measurements [26, 27] and fluorescence lifetime imaging microscopy (FLIM)[16, 28-30] indicate that the fluorescence lifetime of Laurdan depends on the lipid packing. The lifetime in both bands is shorter in $L_\alpha$ than in $S_o$.[31]

The transition dipole moment (TDM) of the chromophore part of Laurdan is perpendicular to the short molecular axis of the naphthalene core. It is generally accepted that the orientation distribution function of the TDM has a maximum along the normal to the membrane and is azimuthally symmetric.[1, 9, 32] Consequently, the fluorescence intensities depend on the orientation of the membrane with respect to the polarization of the excitation light. The corresponding photoselection depends on the lipid packing order.[1, 5, 33] Assuming a rodlike shape for Laurdan, [15] the dependence of *GP* on the photoselection process in giant unilamellar vesicles (GUVs), composed of a single type of lipid, has been interpreted in terms of an intrinsic microheterogeneity of the lipid organization below the microscope resolution.[5, 33-36]

Quantum mechanical and molecular dynamics simulations indicate that in the $L_\alpha$ phase the fluorescent moiety of Laurdan can take orientations near 90º with respect to the membrane normal.[37-43] In this context it is suggested that Laurdan in lipid membranes can take two different conformations which differ in orientation of the carbonyl group with respect to the naphthalene system.[37-40] The carbonyl oxygen points either toward the *β*-position of the naphthalene core (Conf-I) or to the α-position (Conf-II), (Scheme 1). The lipid phase determines the shape of each conformation (see Table 1). In the $L_\alpha$ phase of DPPC, Conf-I takes a bent or L-shape so that the long axis of the naphthalene core forms a substantial angle with hydrocarbon tail of the probe molecule, while Conf-II has the generally assumed elongated shape.[37]

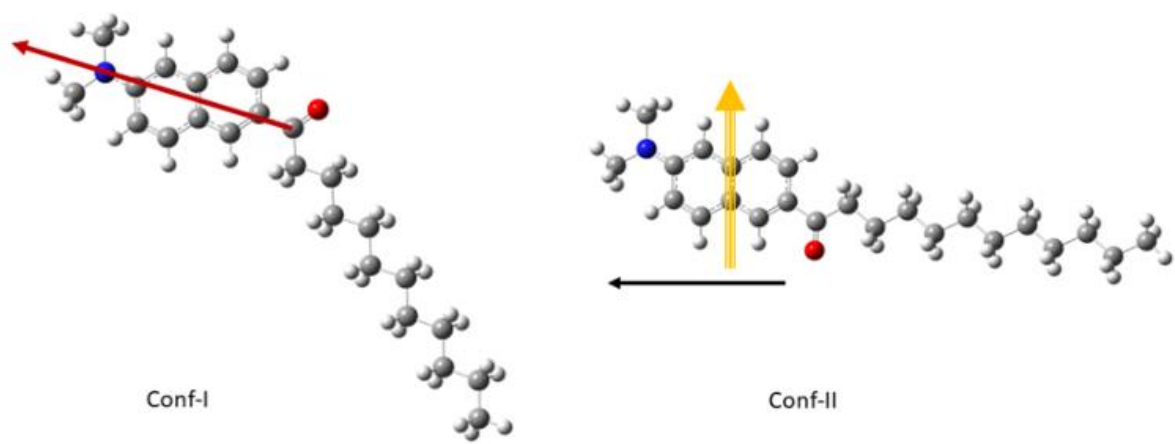

**Scheme 1**. Different conformations of Laurdan in the gas phase with the representation of carbon (grey), oxygen (red), nitrogen (blue) and hydrogen (white) atoms. The transition dipole moment (black), as well as the long (red) and the short (yellow) molecular axes of the headgroup are indicated by vectors.

The opposite is true for DPPC in the $S_o$ phase.[40] The elongated and the L-shape do not interchange in the $S_o$ phase but do so in the $L_α$ phase of DOPC.[39, 40] Although Laurdan moves towards the lipid-water interface during the lifetime of the excited-state,[41] the shape in the excited state is determined by the shape in the ground state.

**Table 1**. The lipid phase of DPPC determines the shape of the conformations of Laurdan

|  | $S_o$ phase | $L_α$ phase |
| --- | --- | --- |
| Conformation-I | elongated | bent |
| Conformation-II | bent | elongated |

The two conformations take different positions in the membrane with the bent shape having the carbonyl oxygen more exposed to the water phase. Given the solvatochromic nature of Laurdan the different accessibility of the two shapes of Laurdan to water molecules has important consequences for the spectral and fluorescence relaxation behavior.

In this work, the relation between GP and the fluorescence lifetime in the blue and green channel is investigated by spectroscopic measurements on macroscopically isotropic solutions of large unilamellar vesicles (LUVs) of DMPC and DPPC as a function of temperature. We performed steady–state and time-resolved fluorescence imaging microscopy of GUVs of DPPC and DOPC at 21ºC as an oriented membrane system using linearly polarized two-photon excitation. The spatial patterns of the measured quantities due to photoselection are discussed in terms of membrane microheterogeneity and the suggested different shapes of Laurdan. Additionally, the relation between *GP* and fluorescence relaxation times in the blue and green channel in each lipid phase is investigated over the optical cross-sections of the GUVs.

**Materials and methods**

**Materials**

Dipalmitoyl phosphatidylcholine (DPPC), dimistroyl phosphatidylcholine (DMPC) were purchased from Avanti Polar Lipids (Alabastre, AL, USA); dioleoyl phosphatidylcholine (DOPC) from Sigma-Aldrich (St. Louis, MO, USA). Laurdan was obtained from Invitrogen/Molecular Probes (Eugene, OR, USA). Dimethylsulfoxide (DMSO), $Na_2HPO_4 \cdot 2H_2O$, $KH_2PO_4$ anhydrous and NaCl were purchased from Sigma-Aldrich and chloroform from Merck (Darmstadt, Germany). Phosphate-buffered saline (PBS; 10 mM, with a 7.4 pH), was used for large unilamellar vesicles (LUVs) preparation.

**Vesicle Preparation and Labelling**

LUVs of DMPC and DPPC were prepared by extrusion. Briefly, the appropriate amount of lipids from the stock solutions prepared in chloroform were mixed and dried under nitrogen flow. The lipid film was then hydrated with PBS heated above the phase transition temperature $T_m$ of the lipids used, and vigorously vortexed, resulting in a suspension of multilamellar vesicles (MLVs). The MLV suspension was subjected to 5 freeze-thaw cycles and extruded (15 times) through a 200 nm polycarbonate membrane using a standard extruder (Avanti Polar Lipids, Alabastre, AL, USA). The extrusion was performed at a temperature above $T_m$ of the lipids used, resulting in a suspension of LUVs with a final lipid concentration of 50 µM. After liposome preparation, Laurdan from a stock solution prepared in DMSO was added into the samples to a final lipid: probe molar ratio of 500:1. Giant unilamellar vesicles (GUVs) with a diameter varying between 10 and 100 µm were prepared by the electroformation method[44] in a home-made closed holder using two square indium tin oxide[45] (ITO) coated coverslips (Praezisions Glas & Optik GmbH, Iserlohn, Germany) separated by a 3 mm rubber spacer. Before each measurement, a fresh solution containing lipids (1.2 mM in chloroform) was mixed with Laurdan (0.6 µM in DMSO) and a few drops (10 µL) were placed on the surface of the bottom ITO-cover glass and the solvent was dried in an oven at 50 °C (15 min). After the chamber was assembled, Milli-Q water was added, at a temperature above $T_m$ of the lipids used, and an AC voltage of 1.5 V amplitude at 10 Hz was applied for 30 min. GUVs images were recorded at room temperature (21-22ºC). Two types of GUVs were prepared: DPPC and DOPC with a probe:lipid ratio of 1:2000.

**Fluorescence Spectroscopy**

Steady-state fluorescence measurements were performed using a FluoroMax 3 spectrofluorimeter (Horiba Jobin Yvon, New Jersey, NJ, USA), equipped with a Peltier-thermostated cell holder. The excitation was at 378 nm and the spectra were recorded in the range of 400 nm to 600 nm with excitation and emission slits set at 3 nm. The spectra were corrected by subtracting the contribution of a liposome suspension without Laurdan. The variation of the spectral sensitivity of the detector was corrected based on the manufacturer file provided in the software. Fluorescence lifetime measurements were performed using a home-made time-resolved fluorimeter based on time-correlated single-photon counting. The excitation light source was a sub-nanosecond pulsed LED head PLS 370 (380 nm, spectral width 15 nm, pulse width 600 ps) controlled with the PDL 800-D driver, both from PicoQuant (Berlin, Germany). The decay curves were recorded using the Time-correlated Single Photon Counting module TimeHarp 200 and a photomultiplier detector PMA182-P-M, also from PicoQuant. The fluorescence decays were measured at 440 nm and at 490 nm by using filters with a band pass of 20 nm (Chroma Technologies, Bellows Falls, VT, USA). The instrument response function (IRF) was obtained through the use of a scattering Ludox solution. The cuvette temperature (5 – 60 ºC) was controlled by a water bath thermostat. Fluorescence decay data were deconvoluted using the FluoFit 100 software package from PicoQuant. The number of counts found in the peak of the curves was higher than 5000. The quality of the fit was judged by the reduced $\chi 2$ value, which takes a value around 1 or slightly larger for a high quality fit.[46]

**Fluorescence Microscopy**

Fluorescence lifetime images (FLIM) were recorded at room temperature (21 ºC) using a confocal laser-scanning microscope (Zeiss LSM 510 META installed on a Zeiss Axiovert 200M; Zeiss, Jena, Germany). Two-photon excitation was obtained using a pulsed titanium-sapphire laser (Mai-Tai Deep-See, Spectra Physics, Newport) tuned at 780 nm. The excitation light was directed to the sample through a water immersion objective from Zeiss (LD C-Apochromat 40x/1.1 W Korr UV-VIS-IR) and separated from the emission with a KP 650 nm dichroic beam splitter from Zeiss. The emission was split into two channels using a dichroic beam splitter at 470 nm, a 405–455 nm filter for the blue channel and a 475–565 nm filter for the green channel. All filters are obtained from Chroma. Two images of 128×128 pixels (with pixel size ranging from 0.6 to 0.28 µm) of optical cross-sections of GUVs were simultaneously collected using two photomultiplier detectors (H7422P-40, Hamamatsu, Japan) connected to a Becker & Hickl SPC830 card (Becker & Hickl GmbH, Berlin, Germany), allowing for time-correlated single photon counting. The intensity in each pixel was obtained by integrating the photons accumulated during the measurement. The IRF was obtained by recording the second harmonic signal obtained from $KH_2PO_4$ crystals. The fluorescence lifetime images were analyzed using SPCImage software from Becker & Hickl. The number of counts found in the peak of the curves was always higher than 2000. The quality of the fit was judged by the reduced $\chi^2$ value at the pixel level. An effective single relaxation time was sufficient to fit the data. The experiments on GUVs were conducted at a laser power below 3.6 mW, as measured on the microscope stage. The equivalence of the lifetime determinations on the spectroscopic setup and the FLIM were verified by control experiments on dyes in solution (p-Terphenyl in ethanol and Rhodamine B in water, respectively). We used a home-made automated polarization controller device to change the polarization of the excitation light. The microscopy body has a slot to insert a slider between the non-descanned dichroic mirror and the scan head output. The slider contains a quarter wave plate to control the ellipticity of the polarization of the light and a half wave plate to rotate the main polarization axis of the illumination beam. Both wave plates can be steered independently. The laser beam first passes the quarter wave plate and then the half wave plate to control the polarization state of the exiting laser beam. The same GUVs were recorded twice, first with linearly polarized light (LP) along the x-axis and thereafter with circularly polarized light (CP).

**Data processing**

Spectra were processed using Origin 8.0 (OriginLab Corporation, Northampton, MA, USA). *GP* values for spectrometric measurements were calculated using equation (1). For microscopy measurements, *GP* values were calculated for each pixel using:

$$GP = \frac{I_{(405-455)} - G I_{(475-565)}}{I_{(405-455)} + G I_{(475-565)}} \quad (2)$$

where $I_{405-455}$ and $I_{475-565}$ are the emission intensities collected in the blue (405 nm – 455 nm) and green channel (475 nm – 565 nm), respectively, and with G the correction factor specific for the experimental setup:[47]

$$G = \frac{I^{DMSO}_{(405-455)}(1 - GP_{theo})}{I^{DMSO}_{(475-565)}(1 + GP_{theo})} \quad (3)$$

where intensities of Laurdan emission in the specific channels were recorded for a sample of Laurdan in DMSO, under the same conditions as for GUVs; $GP_{theo} = 0.207$ and represents the known *GP* value of Laurdan in DMSO at 22 °C.[47]

Below, the quantities observed in the blue channel will be denoted by the subscript 'B' and those in the green channel by the subscript '*G*'. Temperature dependence of *GP* and lifetimes values were fitted in Origin using a Boltzmann sigmoidal function and lipids $T_m$ was estimated from the inflection point. [48, 49]

Images have been processed in MatLab 9b software (MathWorks, Natick, MA, USA) as follows. First an intensity based threshold was used to discard the background pixels and to select only the pixel corresponding to the membrane. *GP* images were processed likewise and the distribution and the average *GP* value over the membrane pixels were obtained. The lifetime images were processed similarly and the distribution and the average relaxation time over the membrane pixels, for each channel were obtained. Finally, the microscopically obtained optical cross-sections of the GUV (Scheme 2) are analyzed in terms of angular plots as follows. The intensities, *GP* and lifetimes are considered as a function of the angle between the polarization of the excitation light and the local membrane normal at the considered position on the optical cross-section of the GUV (Scheme 2). To obtain an acceptable signal-to-noise ratio, averaging over several GUVs is performed. This type of averaging implicitly assumes that heterogeneities are absent at the actual optical resolution in a membrane composed out of a single type of lipid.[5, 33, 35]

When determining the angular dependence of the *GP* and the fluorescence relaxation time with respect to the polarization of the incident light, pixel values within a two degree angular step were binned.

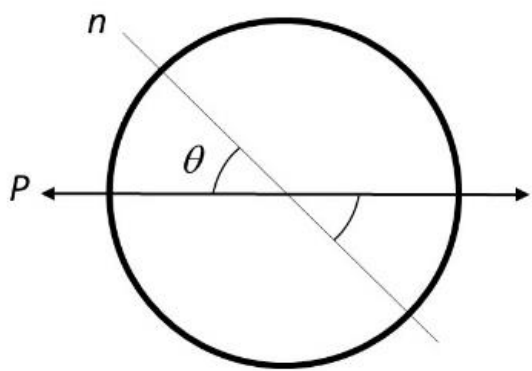

**Scheme 2**. Schematic representation of the optical cross-section of a GUV perpendicular to the optical axis of the microscope. The polarization direction *P* of the excitation light is represented by the double headed arrow. The thin black line shows the local normal *n* to the membrane. The angle between *P* and *n* is denoted by θ. The so-called angular plots in the paper give the observed variable as function of θ. Because of the symmetry of the system the value of θ ranges from 0° to 180°. Note that in general the angle θ is not the angle between *P* and the direction of the excitation TDM of the probe molecule. This TDM is assumed to be azimuthally symmetrically distributed around the local normal to the membrane.

**Molecular dynamics simulations**

An exhaustive description of the computational methodology has been reported previously for DOPC[39] and DPPC[40]. For the convenience of the reader the most important aspects are briefly repeated here. The MD simulations have been performed by means of the Gromacs 5.0.6 software and the GROMOS 43A1-S3 force field.[50, 51] Besides 64 DPPC molecules per layer, the bilayer consisted out of 3314 water molecules along with a physiological concentration of sodium and chlorine counter ions. A pre-equilibrated membrane was used with an energy minimization step of 1 ps and thermalization over tens of ns. One probe molecule was used and the MD run encompassed 400 ns. After 240 ns, the systems were seen to be equilibrated since the orientations of the transition dipole moments showed negligible fluctuations (see Figure S6). All analyses shown here are therefore related to this time window. More complex and realistic cell membranes can be considered, too; they encompass multiple components, phases and phase transitions, which at this stage complicate the interpretation of fluorescence experiments when the interaction of a probe with a more basic environment is not clarified yet. In view of the optical properties of Laurdan, we prefer therefore to focus on the essentials and follow the line of our previous analyses to ensure consistency with earlier results.

Atomic charges were obtained from the Gaussian 09 package of programs, the CAM-B3LYP/cc-pVDZ level of theory and the electrostatic potential (ESP) scheme.[52, 53] The water solvent and lipids were described by means of the TIP3P and Berger parameters, respectively.[54] The integration time step was set to 2 fs, while a cutoff of 1.2 nm was used to compute the Coulomb and van der Waals interactions, within the Particle Mesh Ewald method. [55] The coordinates were saved each picosecond. An orthorhombic box of 5x6x8 nm was used. It was periodic in the x and y directions; the z-direction was normal to the membrane plane. In the framework of a canonical NPT ensemble, the temperature and pressure were set to 298K and 1 bar, respectively. The Nosé-Hoover thermostat and Parrinello-Rahman barostat were used with 5 ps and $4.5·10^{-5}$ $bar^{-1}$ as time constant and compressibility, respectively. [56-58]

**Results and discussion**

**Measurements on LUVs**

Fluorescence emission spectra of Laurdan inserted into DPPC and DMPC LUVs were recorded as a function of temperature (Figure S1.A-C). As expected, the Laurdan emission spectrum shape does not depend on the lipid type in the $S_o$ and $L_α$ phase.[59] Temperature induced membrane changes were assessed by *GP* (Figure 1). LUVs prepared from DMPC or DPPC exhibit *GP* values higher than 0.4 below $T_m$, as is expected for lipid membranes in the $S_o$ phase.[15] The *GP* values decrease with increasing temperature and reach values of about -0.4 for the $L_α$ phase. The temperature dependence of *GP* in DMPC LUV compare well with that obtained by others.[60] In Table 2 are reported the $T_m$ values estimated both LUV systems and compared with thermodynamically obtained values reported in the literature.[61] Near the inflection point the temperature dependence of *GP* in DPPC is more pronounced than for DMPC. The different temperature dependence of *GP* near $T_m$ for DPPC and DMPC has been reported in the literature.[62-64] The values of $T_m$ found in the literature for DMPC LUV show more variation than those for DPPC LUV.

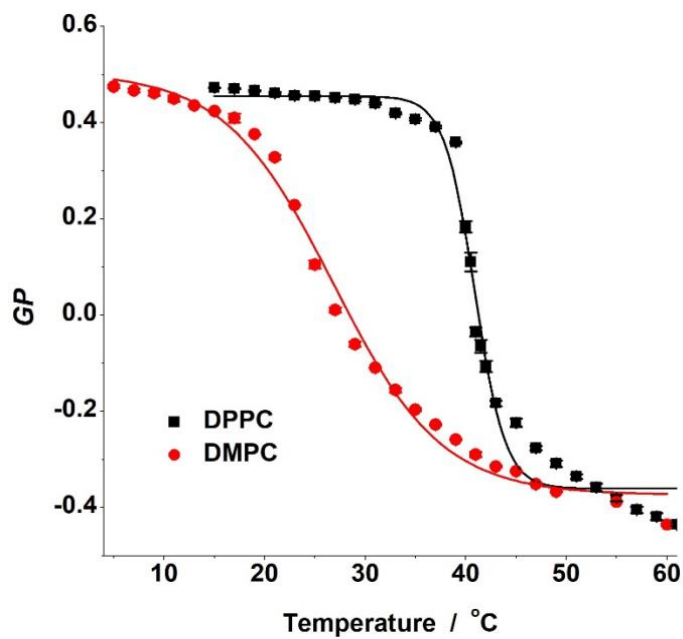

**Figure 1.** Variation of GP values with temperature for LUVs prepared from DMPC and DPPC. The error bars correspond with the standard deviation resulting from three repeats. The solid lines represent the sigmoidal fit used to obtain the $T_m$ values.

**Table 2.** Transition temperatures (in °C) obtained by analyzing the temperature dependence of GP and the fluorescence relaxation times of Laurdan in LUV.

|  | DMPC | DPPC |
|---|---|---|
| *GP* | $26.8 \pm 0.5$[a] | $40.9 \pm 0.2$ |
| $\tau_{blue}$ | $24.1 \pm 0.7$ | $40.7 \pm 0.2$ |
| $\tau_{green}$ | $30.8 \pm 1.3$ | $41.1 \pm 0.4$ |
| [*Ref. [61]*] | $22.2 \pm 2.0$ | $41.4 \pm 0.1$ |

[a] Uncertainties are standard errors obtained from the fitting parameters.

The fluorescence lifetime of Laurdan was recorded in the blue (440 nm) and the green (490 nm) channel at different temperatures for each type of vesicle composition (Figure 2). In our LUV experiments the fluorescence decays were well described ($\chi^2 < 1.2$) by a single exponentially decaying function in both the blue and green channel at the considered temperatures. The relaxation times in both the blue ($\tau_B$) and green emission band ($\tau_G$) are larger in the $S_o$ phase than in the $L_\alpha$ phase. The probe environment is more accessible to water molecules in the $L_\alpha$ phase and this leads to shorter lifetimes in the blue band.[33, 65] In each lipid phase the relaxation time in the green band is longer than in the blue band. The emission in the green channel is related to the molecular rearrangements near the enhanced dipole moment of Laurdan upon excitation.[33, 65] A faster rearrangement is possible in the $L_\alpha$ phase so that a shorter lifetime is obtained as compared with the $S_o$ phase.

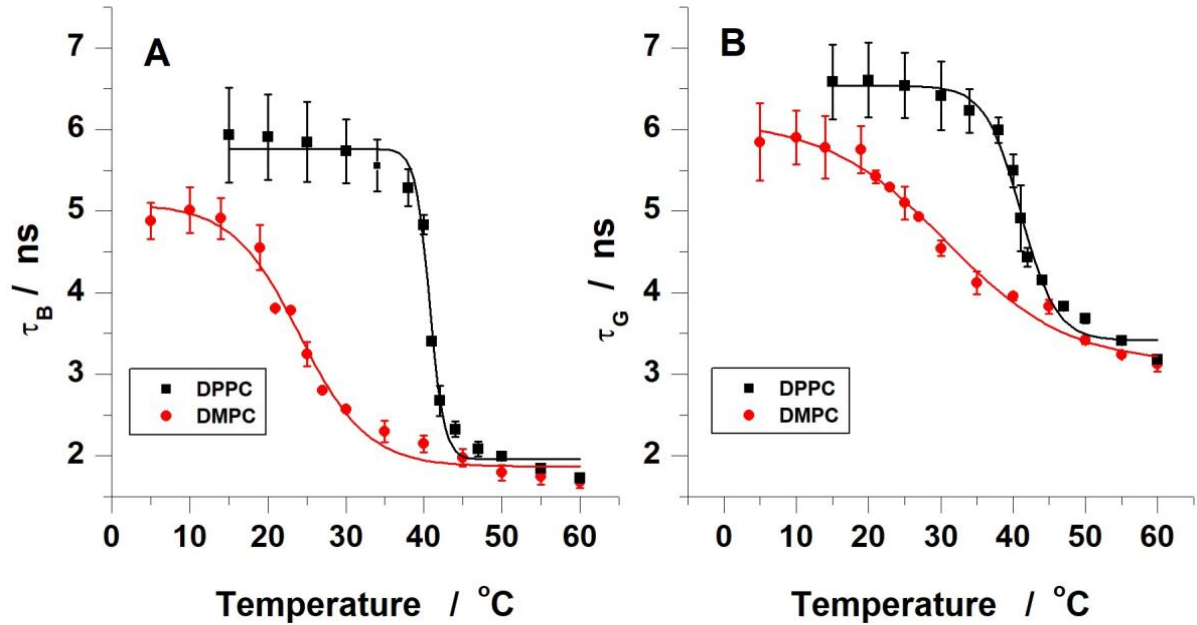

**Figure 2.** Fluorescence relaxation times of Laurdan in the membrane of LUVs recorded in blue (A) and green (B) channels. The solid lines represent the sigmoidal fit used to obtain the Tm values. Error bars indicate the standard deviations resulting from three repeats.

DMPC LUVs in the $S_o$ phase yields smaller relaxation times in comparison to DPPC LUVs. In the blue channel, a fluorescence lifetime of 4.9 ± 0.2 ns was recorded for DMPC and 5.9 ± 0.6 ns for DPPC, while in green channel the relaxation times were 5.8 ± 0.5 ns and 6.6± 0.5 ns, respectively. The somewhat shorter lifetimes in DMPC in the $S_o$ phase seem to suggest that the Laurdan molecules inserted in DMPC in the $S_o$ phase are differently exposed to the water interface. However, in the $L_\alpha$ phase the observed relaxation times were essentially independent of the lipid chain length, and take values of 1.7 ± 0.1 ns and 1.7 ± 0.1 ns in the blue channel and 3.1 ± 0.1 ns and 3.18 ± 0.04 ns in the green channel. Vequi-Suplicy *et al.*[66] using a bi-exponential, global analysis over various emission wavelengths (470, 480 and 490 nm) for Laurdan found for DMPC LUV lifetimes of ~ 6 ns and ~8 ns in the $S_o$ phase and lifetimes of ~ 2 ns and ~ 4 ns in the $L_\alpha$ phase. The pre-exponential factors were positive throughout; below the phase transition the shorter lifetime contributed mostly while the longer lifetime dominated above the phase transition.[66] The lifetimes we observed for DPPC LUV are comparable with those reported by Watanabe *et al.*[65]

The temperature dependence of the fluorescence lifetimes in the two channels for the LUVs displays also a sigmoidal temperature dependence and the same procedure as for *GP* was followed to provide estimates for the transition temperature $T_m$ (Table 2). The estimates for $T_m$ obtained for DPPC in both emission bands are in very good agreement with the literature value. The results obtained for DMPC deviate to some extent, especially the high value resulting for $T_m$ from the temperature dependence of the lifetimes in the green channel. The temperature variation of the relaxation times observed in the green channel for DMPC is less pronounced as compared to that of the other lifetimes so that a larger uncertainty on the estimation of $T_m$ can be expected. However, this alone cannot explain the large deviation from the literature value. It must be concluded that in DMPC the factors involved in the reorganization of the environment around the excited Laurdan molecule that determine $\tau_G$ have a different temperature dependence as compared to the main phase change of DMPC.

Figure 3A displays $\tau_G$ versus $\tau_B$ for the DMPC and DPPC LUV systems over the considered temperature range. The two lifetimes are strongly correlated. Two different regions in the correlation plots can be identified. The correlation of lifetime values recorded at temperatures above $T_m$ (smaller lifetimes) essentially overlaps between DMPC and DPPC, indicating that in the fluid phase Laurdan is sensing a similar environment for a given lifetime in one of the two channels. The correlation for the larger lifetimes in the $S_o$ phase is also pronounced for each lipid type and exhibits a similar trend for DMPC and DPPC. However, the uncertainty on the position of the dots in the correlation plot is larger.

Figure 3B shows the relation between *GP* values (Figure 1) and the relaxation times in the blue and in the green channel (Figure 2). For a given *GP* value the relaxation time in the green channel is always larger than the relaxation time in the blue channel because of the slow solvent relaxation near the phospholipid headgroup.[22] The lifetimes in both channels are highly and positively correlated with *GP* over the range -0.4 to 0.6. For the $L_\alpha$ phase the correlation between *GP* and the lifetimes is almost identical for DMPC and DPPC.

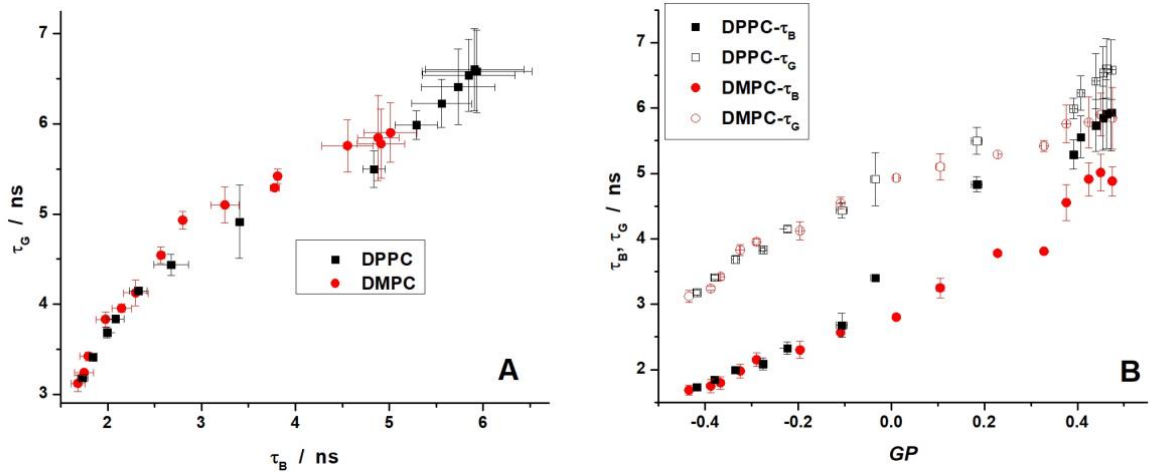

**Figure 3.** (A) $\tau_G$ versus $\tau_B$ for DMPC and DPPC and LUVs at different temperatures. (B) $\tau_G$ and $\tau_B$ (Figure 2) versus *GP* values (Figure 1) at corresponding temperatures. For each condition at least three independent repeats were performed.

The observed positive correlations can be rationalized as follows. The lipid phase determines the accessibility of the excited fluorescent probe to water molecules and the relaxation rate of the surrounding water molecules. In the So phase the excited molecules stay largely in the non-relaxed form. The solvent relaxation rate is very small so that the lifetime of the excited molecule observed in the blue channel is correspondingly long and the emission is predominantly in the blue channel. This leads to a high positive *GP* value. The lifetime in the green channel is even longer than in the blue channel because the relaxed form finds its origin in the nonrelaxed form. In the $L_\alpha$ phase the excited probe is more exposed to water and the solvent relaxation rate is relatively fast. This facilitates a faster conversion of the excited molecules from the nonrelaxed to the relaxed form. This results in a shortening of the lifetime observed in the blue channel because of the quenching by water molecules and the quicker conversion from the nonrelaxed to the relaxed form.[33, 65] The latter yields a strong emission in the green channel yielding a negative *GP* value. The lifetime in the green channel is also short but somewhat longer than the lifetime in the blue channel as explained above. It can be concluded that both the number of water molecules, i.e. polarity, and their relaxation rate increase upon increasing the temperature for DMPC and DPPC LUVs.

It can be stated that at temperatures above $T_m$ the behavior of Laurdan in DMPC and DPPC is very similar but that at $T_m$ and at lower temperatures the difference in chain length between DMPC and DPPC leads to a dissimilar behavior of Laurdan.

**Microfluorimetry of GUVs under linearly polarized excitation: intensity and GP**

The spatial variation of the photoselection effect under linearly polarized excitation light on the fluorescence intensities, *GP* and the fluorescence relaxation times in the blue and green band are investigated in GUVs under a microscope. The use of two-photon excitation allows for a more pronounced photoselection effect as compared to one-photon excitation. The observed quantities at each position along the optical cross-section of the GUV (Scheme 2) are represented in so-called angular plots.

Microfluorimetric measurements were performed at room temperature on GUVs that exhibit different phases: $S_o$ phase (DPPC) and $L_\alpha$ phase (DOPC). Representative images of the fluorescence intensities and the fluorescence relaxation times in the blue and green emission band, and the calculated *GP* obtained under linearly polarized excitation light and their control experiments with circularly polarized excitation light are shown in Figure S2 for DPPC and in Figure S3 for DOPC. When the excitation light is linearly polarized, a photoselection effect is observed in the fluorescence intensity images. As expected, this effect is stronger when the lipids are in the $S_o$ phase in comparison to the $L_\alpha$ phase, where the intensity is more evenly distributed. [5, 33]

Figure 4 shows the angular variation of *GP* and the normalized fluorescence intensities in the two channels for Laurdan in DPPC and DOPC under linearly polarized excitation. The fact that the angular plots show a distinct pattern validates the assumption concerning the homogeneity of the lipid membrane at the optical resolution. Given that the photoselection probability depends on the angle between the polarization direction and the TDM of the molecule in the ground state and the symmetry of the optical cross-section of the GUVs, the angular plots are symmetric around 90º.

The high values of *GP* in DPPC are fairly constant over a large angular range and exhibit only a small angular variation in the region around 90º (Fig. 4A). As expected, much lower *GP* values are observed for DOPC. The values decrease continuously towards 90º over a wide angular range.

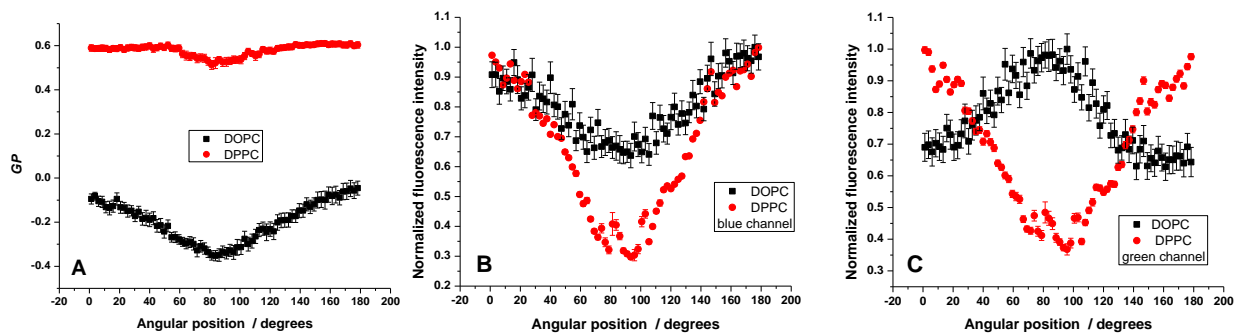

**Figure 4.** *GP* (A) and normalized fluorescence intensities in the blue (B) and green (C) channels versus the angle between the orientation of the polarization of the excitation light and the local normal to the membrane for GUVs of DOPC and DPPC at room temperature. The angular position of 90° is perpendicular to the linear polarization of the incident light. The image was divided in 150 equal angular regions and for the pixels found in each region was calculated the average intensity and GP values. Data are reported as mean ± standard error of the mean [52] for 15 DOPC GUVs and 27 DPPC GUVs. Representative experiments are presented in Figures S2 and S3.

Control experiments with circularly polarized light do not exhibit an angular variation of *GP* (Figure S4).

Because *GP* is a ratio of two intensities at different emission bands the actual number of excited molecules is not relevant. Therefore, *GP* for a homogeneous membrane is independent of the angular position on the cross-section of the GUV. However, the angular variation of *GP* in the $L_\alpha$ phase exhibits a pronounced angular dependence. The highest *GP* values are observed in the direction parallel to the polarization of the excitation light while the lowest *GP* values are situated in a direction perpendicular to the polarization. This observation has also been made by Bagatolli *et al.*[35] for the cross-section of individual GUVs at and below the phase transition. Taking that Laurdan behaves as a stiff rod with the TDM along the long axis of the rod which is oriented along the fatty acid chains of the phospholipid[35], the angular dependence of *GP* in the $L_\alpha$ phase has been modelled in terms of microheterogeneity at a scale below the size of a pixel.[5, 33, 35] In this interpretation Laurdan in the more gel-like microdomains is mainly excited along the direction of the polarization direction of the excitation light (higher *GP*), while liquid-crystal microdomains, where the TDM of Laurdan can make large angles with the membrane normal, manifest themselves clearly by excitation at 90° (lower *GP*). At intermediate angles a weighted contribution of the two types of microdomains has to be considered. Our angular plots of *GP* are obtained by averaging optical cross-sections of multiple GUVs. This suggests that the spatial scale of the assumed microheterogeneity must be very small indeed as an angle dependent average *GP* is registered at the pixel level.

Previous studies[18, 35] made the observation that *GP* in the gel state is essentially independent on the angular position on the cross-section and concluded that the intrinsic microheterogeneity is not present in the $S_o$ phase. However, we observe a small angular variation near 90º. By application of the spectral phasor analysis to imaging microscopy of GUV under linearly polarized two-photon excitation of Laurdan, Golfetto *et al.* indicated that the region near the equator (0º/180º) and the region at the poles (90º) have distinct spectral properties.[18] For both the $L_\alpha$ and $S_o$ phase the region at the poles is characterized by a different phasor as compared to the region at the equator. The region with a distinct phasor at the poles is very narrow for the $S_o$ phase. As the spectral phasor correlates with *GP*, our results for both DOPC and DPPC are in full agreement with these observations. This suggests that there is also microheterogeneity in the $S_o$ phase. It is somewhat remarkable that *GP* in the $S_o$ phase can be determined at 90º. This assumes that the TDM of the rodlike probe can make large angles with respect to the membrane normal as the photoselection under two-photon excitation leads to a high angular photoselectivity. This is unexpected for the $S_o$ phase.

To further explore the pronounced angular dependence of GP in DOPC it is interesting to look at the ratio of $I_G / I_B$ in more detail. Evidently, the ratio of the fluorescence intensities in the blue and green channel, $I_B$ and $I_G$, changes substantially. At 0º/180º *GP* ≈ -0.10 implying that $I_G / I_B$ ≈ 1.2, while at 90º *GP* ≈-0.35 with a corresponding ratio $I_G / I_B$ ≈ 2.

It is also interesting to look at the angular dependence of $I_B$ and $I_G$ separately. To obtain an acceptable signal-to-noise ratio averaging over several GUVs is used by summing the angular intensity plots that are normalized their maximum value. The resulting angular plots for the blue and green channel for DPPC and DOPC are shown in Figure 4B and C.

For both DOPC and DPPC the normalized fluorescence intensity in the blue channel decreases towards 90º. This angular variation is, as expected for a rodlike shape of Laurdan with the absorption TDM essentially along the axis of the molecule, with an orientation distribution function with a maximum along the normal to the membrane. The variation of the normalized intensity in the blue channel for DPPC ($S_o$ phase) is more pronounced as compared to DOPC ($L_\alpha$ phase) as has been observed by others.[35]

This can be understood as the orientation distribution function of Laurdan in the $S_o$ phase is definitely more narrow than in the $L_\alpha$ phase. Gasecka *et al.*[32] performed polarized two-photon excitation experiments of Laurdan in mixed lipids GUV exhibiting both the liquid ordered and disordered lipid phase. No specific spectral filtering was used at the emission side. Assuming a rodlike shape for Laurdan they found that the orientation distribution can be described by a Gaussian function with a symmetry axis along the normal to the membrane. The half width at half maximum of the distribution function is $(36 \pm 5)°$ in the liquid ordered phase and $(71 \pm 4)°$ in the liquid disordered phase.[32] The angular variation of the normalized intensity for DPPC in the green channel is similar to that in the blue channel. This suggests that for DPPC the molecules emitting in the blue or the green channel have similar orientation distribution functions with a maximum along the membrane normal. Surprisingly, the angular plot of the intensity in the green channel for DOPC is almost a mirror image of that in the blue channel and reaches a maximum at 90°. A broad orientation distribution function with a maximum along the normal to the membrane, as can be deduced from the normalized intensity profile in the blue channel, cannot explain the maximum at 90° in the green channel. A convex angular *GP* pattern can be obtained by a convex angular pattern in both the blue and the green channel. The actual data for Laurdan in DOPC show that the observed convex angular *GP* pattern results from a convex angular pattern in the blue channel combined with a concave angular pattern in the green channel.

Given that the photoselection occurs in the ground state, the new observation of the angular pattern in the green channel suggests that the molecules emitting either in the blue or the green channel must have different orientation distribution functions in the ground state. The maximum intensity we observe in the green channel at 90° implies for the intrinsic spatial microheterogeneity model that the elongated form of Laurdan in this angular region must be essentially parallel to the membrane-water interface of the liquid-crystal microdomains. It is however very unlikely that the elongated Laurdan molecule will be oriented like this because of its hydrophobic moiety. The partial charges on its fluorescent moiety make it evenly unlikely that Laurdan positions in the middle of the lipid membrane, as has been reported for a fraction of the rod-like membrane probe 1,6-diphenyl-1,3,5-hexatriene.[67],[68] The orientation distribution function corresponding with the angular intensity profile in the green channel reaches a maximum value in a plane perpendicular the normal to the membrane. In a first approximation this can denoted as an- outside-a-cone model.[69]

The observed angular variation of the intensity in the green channel for DOPC can be interpreted in terms of a bent or L-shape of Laurdan, a conformation suggested by molecular dynamics calculations[37, 39-42] in the $L_\alpha$ phase, as well as an elongated rodlike shape.

Based on experimental observations, a bent shape of Laurdan has been suggested also upon incorporation in giant liposomes composed of the polar lipid fraction from the thermoacidophilic archaebacterial Sulfolobus acidocaldarius[70] and in the phospholipid shell of microbubbles.[71]

In the $L_\alpha$ phase the L-shape is taken by Conf-I and the elongated shape by Conf-II (Table 1). The different angles between the direction of the excitation polarization and the local membrane normal determine the relative contribution of the elongated and the bent shape of Laurdan. The relative contribution of the L-shape of Laurdan to the fluorescence intensity in the green channel for DOPC increases with the angle between the excitation polarization and the membrane normal. Theoretical calculations indicate that the L-shape is more exposed to water molecules than the elongated shape.[37, 40] Therefore, the L-shape can be expected to contribute more in the green channel than the elongated shape.

The presence of the L-shape of Laurdan in the $L_\alpha$ phase necessitates a closer look at the interpretation of the intensity variation in the blue channel for DOPC. When the angle between the excitation polarization and the local membrane normal changes from 0° or 180° towards 90°, the relative excitation probability of the L-shape increases. However, as a minimal value of the normalized intensity in the blue channel is reached at 90°, the intensity contribution of the L-shape in the blue channel must be smaller than that of the elongated form. This interpretation is consistent with the stronger contribution of the L-shape in the green channel at 90° because of the different exposure to water.

In this new interpretation, the elongated and L-shape of Laurdan in the $L_\alpha$ phase contribute differently in the two emission channels. The angular decrease of the intensity in the blue channel and the angular increase of the intensity in the green channel indicate that the elongated shape dominates in the blue channel and that the bent form contributes mostly in the green channel. The angular profile of *GP* in DOPC can then be interpreted in terms of the angle dependent relative contribution of the elongated and bent conformation of Laurdan. Rather than the elongated form, the bent one can be associated with a lower *GP* as the latter one is more exposed to water molecules.

The angular intensity profiles in the blue and green channel for DPPC near 90° are rather broad and do not reach very low values. This angular dependence of the intensity can be compared with results reported in a similar polarization study with two-photon excitation of the rodlike probe DPH in a rigid lipid membrane.[72] In the latter case the normalized intensity profile reaches lower relative values as reported here for Laurdan. This suggest the possible presence of the bent form of Laurdan. However, the prevalence must be rather low. Molecular dynamics calculations suggested the possibility for a bent conformation of Laurdan also in the $S_o$ phase.[40] Here, the results of that work have been explored further. In the $S_o$ phase the L-shape is taken by Conf-II and the elongated shape by Conf-I (Table 1). Figure 5 shows for the elongated form (Conf-I) and the bent form (Conf-II) the correlation between the angle of the TDM in the ground state with the membrane normal and the depth in the membrane as measured by the distance of the carbonyl oxygen of Laurdan with respect to the averaged position of the phosphor atoms of the lipids. We would like to note to the reader that the horizontal axis is different from the one which we used in many of our previous publications, where we mainly focused on the distance to the membrane center. [38-40, 49]

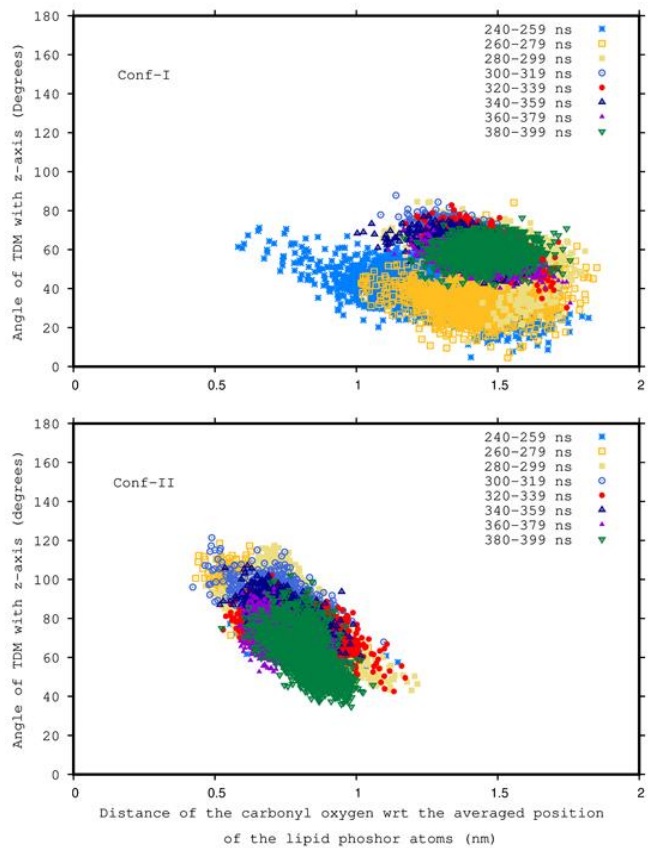

**Figure 5**. Molecular dynamics simulations of Laurdan in DPPC at room temperature. On the vertical axis, the angle of the TDM with the z-axis is given, while on the horizontal axis the distance of the carbonyl oxygen has been typeset with respect to the averaged position of the phosphor atoms of the lipids. Conf I takes the elongated shape. Conf II takes the bent shape. The considered Z-axis is along the membrane normal. Time window of the displayed data is 240 ns – 400 ns. For both conformers of Laurdan, the TDM has been denoted as a vector, which points from the carbonyl carbon to the amino group (see Scheme 1).

From these simulations it can be concluded that when Laurdan takes positions deeper in the membrane, the angle between the TDM and the membrane normal decreases for both conformations. In this way the lipid packing is less disturbed. Only the L-shape (Conf-II) can be found at lower depths allowing for a larger angle between the transition moment and the membrane normal. The L-shape (Conf-II) is located at the more polar region of the membrane leading to a lower *GP* with respect to the elongated form (Conf-I). As the photoselection near 90º enhances the contribution of the L-shape, one can expect a lower *GP* at 90º. This corroborates the observed angular *GP* plot for DPPC.

**Microfluorimetry of GUVs under linearly polarized excitation: fluorescence lifetimes**

Time-resolved fluorescence experiments with linearly polarized light were performed on DPPC and DOPC GUV at room temperature to explore the angular variation of the effective fluorescence lifetimes of Laurdan. The angular dependence of the effective lifetimes, $\tau_B$ and $\tau_G$, for DOPC and DPPC are shown in Figure 6. The lifetimes in the blue channel are shorter than in the green channel as the emission in the green channel assumes a solvent rearrangement near the excited molecule. The lifetimes in DPPC and DOPC exhibit an angular dependence in the blue and green channel, apart from the short lifetime in the blue channel for DOPC that is essentially constant within the time resolution of the instrument (Fig. 6B). The lifetime value does not depend on the number of excited molecules so that just the number of photoselected molecules cannot explain the angular dependence.

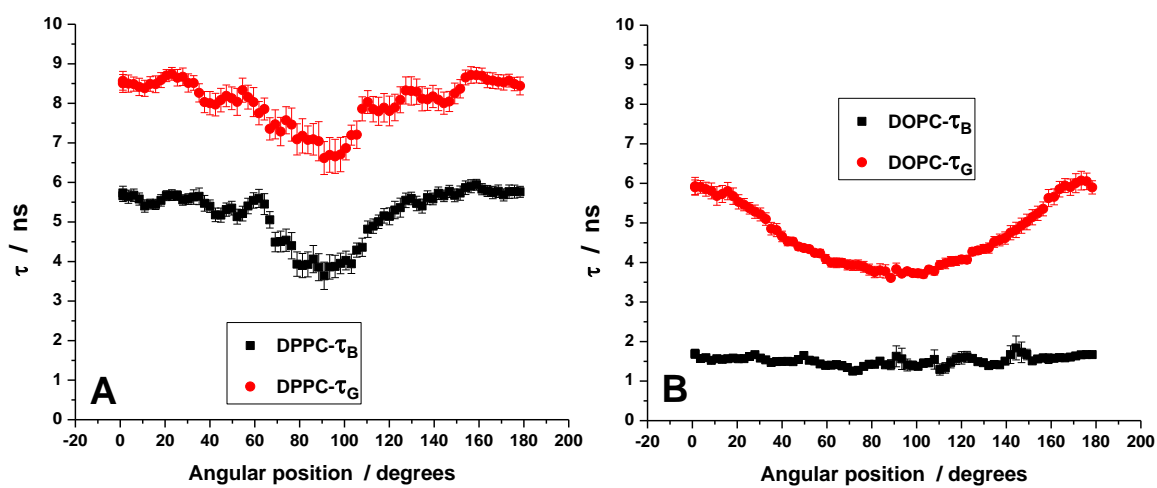

**Figure 6.** Fluorescence lifetimes of Laurdan recorded in blue and green channel versus the angle between the orientation of the polarization of the excitation light and the local normal to the membrane for GUVs of DOPC and DPPC at room temperature. The angular position of 90º is perpendicular to the linear polarization of the incident light. The image was divided in 150 equal angular regions. Data are reported as mean ± standard error of the mean [52] for 15 DOPC GUVs and 27 DPPC GUVs. Representative experiments are presented in Figures S2 and S3.

For the $S_o$ phase of DPPC the angular dependence in the lifetimes in both emission channels exhibit a similar pattern: a rather narrow angular region with lower values that reach a minimum at 90º. The similarity of the angular patterns points to a positive correlation between $\tau_B$ and $\tau_G$ (Figure 7A). The angular region 60º – 120º where the smaller lifetimes are observed is also the region with the lower *GP*. The angular dependence of the lifetime in DPPC is unexpected under the assumption of a homogeneous lipid phase. The angular dependence of the lifetimes can be interpreted in terms of the different conformations as discussed for the angular plots of *GP* and the fluorescence intensities in the blue and green channel. Within the context of this model, the observed effective lifetime at any angle is then a weighted average of the lifetimes of the different conformations. As the orientation distributions in the gel phase are rather narrow, it can be expected that essentially only the bent shape contributes in the angular region at 90º with the lower lifetime values. As discussed above, the molecules with a bent shape are closer to the water interface than the elongated shape, corresponding with a lower *GP*. This leads to lower lifetime in the blue channel because of quenching by water molecules. The lower value of the bent form in the green channel can be rationalized in terms of a somewhat faster solvent relaxation as compared to the elongated shape.

As can be expected for the $L_\alpha$ phase of DOPC, the lifetimes in the two emission bands are smaller than in the $S_o$ phase because of enhanced penetration of water molecules in the bilayer and faster solvent relaxation. The angular dependence of the lifetime in the green channel shows here a variation over a much broader angular range, similar to that of *GP*, while the shorter fluorescence lifetime in the blue channel is essentially independent of the angular position on the optical cross-section of the GUV. In the $L_\alpha$ phase the orientation distribution functions of the two conformations of Laurdan can be expected to be much wider than in the $S_o$ phase so that these overlap. This can explain the gradual changes of the lifetime in the green channel. The more exposed bent shape undergoes a faster solvent relaxation leading to a shorter lifetime in the green channel. As indicated by the angular intensity variation in the green channel, the contribution of the L-shape increases towards 90º where the lifetime reaches its smallest value. Apparently, the different axial positions of the two conformations do not lead to a noticeable angular variation in the lifetime in the blue channel. Control experiments with circularly polarized light did not reveal any angular bias for the lifetimes (Figure S5).

Comparison of the results obtained under LP and CP excitation can provide information on the relative contribution of the conformations of Laurdan in the $L_\alpha$ and $S_o$ phases. The average CP value of *GP* ($\approx$ 0.6) for DPPC (Figure S4) is essentially determined by the elongated shape because of the contribution by the bent shape is restricted to a narrow angular range. This is in contrast with the situation for DOPC. The average CP value of *GP* ($\approx$ -0.2) (Figure S4) is essentially the weighted average of the LP *GP* values obtained at 0º/180º and at 90º, signifying that the contribution of the bent form is substantial. The small contribution of the bent form in DPPC and its significant contribution in DOPC is confirmed by the comparison of the lifetime's values in the blue and green emission bands for DPPC and DOPC. The average CP values for the lifetimes in the blue channel are $\approx$ 5.5 ns (DPPC) and $\approx$ 1.5 ns (DOPC), and in the green band $\approx$ 9 ns (DPPC) and $\approx$ 4.5 ns (DOPC) (Figure S5). Again, these values for DPPC are close to those obtained under LP at 0º/180º, while for DOPC the CP value for the lifetime in the green band is more to the LP value at 90º than to the value at 0º/180º.

**Microfluorimetry of GUVs under linearly polarized excitation: correlation between *GP* and fluorescence lifetimes**

The correlation plots between *GP* and fluorescence lifetimes $\tau_B$ and $\tau_G$ as well as the correlation between $\tau_B$ and $\tau_G$ corresponding to the data of Figure 4 and Figure 6, are shown in Figure 7. The lifetimes $\tau_B$ and $\tau_G$ recorded in DPPC GUVs (Fig. 7A) are strongly correlated. The correlation plot exhibits a similar tendency as for LUVs (Figure 3A). $\tau_B$ recorded for DOPC GUVs has a narrow distribution while the spread of $\tau_G$ is much wider. Nevertheless, the correlation plot for $\tau_B$ and $\tau_G$ in DOPC (Figure 7A) exhibits a definite positive slope.

The correlation of GP with $\tau_B$ and with $\tau_G$ in DOPC and DPPC GUVs is shown in Figure 7B. The tendency of the correlations is similar to that obtained for LUVs (Figure 3B). The similarity of the correlations obtained for LUVs due to temperature changes and those obtained for GUVs by exploring the angular plots due to photoselection indicate a rather general nature of these correlations. However, where changes in temperature lead to changes in membrane organization (LUV), the photoselection that underlies the angular plots provide evidence for different conformations of the membrane reporter molecule Laurdan (GUV).

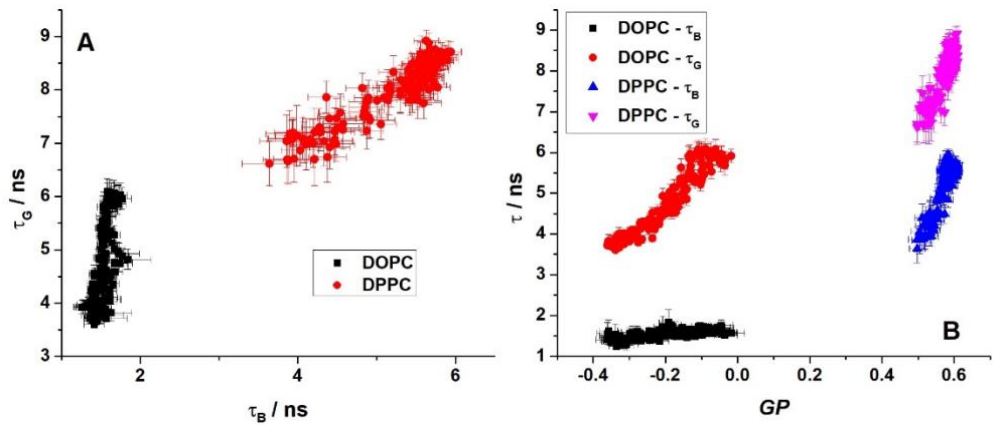

**Figure 7.** Correlation between the two lifetimes (A) and lifetimes and *GP* (B) for Laurdan in DOPC and DPPC GUVs recorded with linearly polarized excitation light. Experiments were performed at room temperature.

The narrow distribution of *GP* in $S_o$ and the broad distribution in $L_\alpha$ is apparent. The difference in the width of the distributions of *GP* of the various phases can be partially ascribed to the intrinsic dependence of *GP* on the ratio $R = I_B/I_G$. A small change in the ratio $R$ has more effect on the value of $GP$ at low $GP$ values ($L_\alpha$) than at higher $GP$ values ($S_o$).

**Conclusion**

In the present study we investigated by both spectroscopic and microscopic techniques the correlation between $GP$ values and the fluorescence lifetimes recorded in the blue and green emission band of Laurdan embedded in vesicles with different lipid organization. For both techniques, we found a strong and positive correlation between the fluorescence lifetimes and the $GP$ values. The fluid membrane ($L_\alpha$) is characterized by shorter fluorescence lifetimes and smaller $GP$ values, while the rigid membrane ($S_o$) is characterized by longer lifetimes and larger $GP$ values. The angular dependence of the $GP$ values and fluorescence lifetimes observed in microfluorimetric experiments under linearly polarized excitation is rationalized by considering two shapes for Laurdan. The elongated shape is characterized by longer lifetimes and larger $GP$ values, while the L-shape is more exposed to water, yielding shorter lifetimes as well as smaller $GP$ values. To our knowledge, this is the first time that experimental data obtained with Laurdan in lipid bilayers suggest that Laurdan can take a bend shape in addition to the commonly accepted rodlike shape, as already reported in computational studies.

**CRediT authorship contribution statement**

**Mihaela Bacalum**: Data curation, Formal analysis, Investigation, Methodology, Visualization, Writing – original draft, Writing – review & editing; **Mihai Radu**: Funding acquisition, Supervision, Resources, Writing – review & editing; **Silvio Osella**: Formal analysis, Investigation, Methodology, Resources, Software, Visualization, Writing – review & editing; **Stefan Knippenberg**: Formal analysis, Investigation, Methodology, Resources, Software, Supervision, Visualization, Writing – review & editing; **Marcel Ameloot**: Conceptualization, Formal analysis, Funding acquisition, Project administration, Supervision, Writing – original draft, Writing – review & editing;

**Declaration of Competing Interest**

The authors declare that they have no known competing financial interests or personal relationships that could have appeared to influence the work reported in this paper.

## Data availability

Data will be made available on request.

## Acknowledgements

The authors are grateful to prof. L. Bagatolli for discussions at some point of this work, to Dr. N. Smisdom for advices regarding the use of the confocal microscope and data analysis of the confocal images and to Dr. R. Paesen for the home-made automated polarization controller device. The authors thank prof. M. Roeffaers for the ITO coated coverslips that were essential in the generation of GUV. S. O. is grateful to the National Science Centre, Poland for funding (grant no. UMO-2018/31/D/ST4/01475 and UMO/2020/39/I/ST4/01446). Computational time was at the Polish side provided by the Interdisciplinary Centre for Mathematical and Computational Modelling at the University of Warsaw (ICM UW) under grants no. G83-28 and GB80-24, while in Belgium the Flemish Supercomputer Centre (VSC) and the Herculesstichting are acknowledged.

## References

1. Bagatolli, L.A., *To see or not to see: lateral organization of biological membranes and fluorescence microscopy.* Biochimica et Biophysica Acta, 2006. **1758**(10): p. 1541-56.
2. Gunther, G., et al., *LAURDAN since Weber: The Quest for Visualizing Membrane Heterogeneity.* Acc Chem Res, 2021. **54**(4): p. 976-987.
3. Parasassi, T., et al., *Quantitation of lipid phases in phospholipid vesicles by the generalized polarization of Laurdan fluorescence.* Biophys J, 1991. **60**(1): p. 179-89.
4. Bagatolli, L.A., et al., *A model for the interaction of 6-lauroyl-2-(N,N-dimethylamino)naphthalene with lipid environments: implications for spectral properties.* Photochem Photobiol, 1999. **70**(4): p. 557-64.
5. Bagatolli, L.A. and E. Gratton, *Two photon fluorescence microscopy of coexisting lipid domains in giant unilamellar vesicles of binary phospholipid mixtures.* Biophys J, 2000. **78**(1): p. 290-305.
6. Bagatolli, L.A., *Monitoring Membrane Hydration with 2-(Dimethylamino)-6-Acylnaphtalenes Fluorescent Probes.* Subcell Biochem, 2015. **71**: p. 105-25.
7. Bagatolli, L.A., *Laurdan Fluorescence Properties in Membranes: A Journey from the Fluorometer to the Microscope.* Fluorescent Methods to Study Biological Membranes, ed. Y.D. Mély, G. Vol. 13. 2012, Heidelberg: Springer: Berlin, . 3-36.
8. Jurkiewicz, P., et al., *Headgroup hydration and mobility of DOTAP/DOPC bilayers: a fluorescence solvent relaxation study.* Langmuir, 2006. **22**(21): p. 8741-9.
9. Sanchez, S.A., et al., *Laurdan Generalized Polarization: from cuvette to microscope*, in *Modern research and educational topics in microscopy: applications in physical/chemical sciences*, Méndez-Vilas and J. A.; Díaz, Editors. 2007, Formatex Research Center. p. 1007.
10. Marini, A., et al., *What is solvatochromism?* J Phys Chem B, 2010. **114**(51): p. 17128-35.
11. Tomin, V.I., M. Brozis, and J. Heldt, *The red-edge effects in Laurdan solutions.* Z. Naturforsch. , 2003. **58a**: p. 109.
12. Jurkiewicz, P., et al., *Lipid hydration and mobility: an interplay between fluorescence solvent relaxation experiments and molecular dynamics simulations.* Biochimie, 2012. **94**(1): p. 26-32.
13. Parasassi, T., et al., *Phase fluctuation in phospholipid membranes revealed by Laurdan fluorescence.* Biophys J, 1990. **57**(6): p. 1179-86.
14. Parasassi, T., et al., *Influence of cholesterol on phospholipid bilayers phase domains as detected by Laurdan fluorescence.* Biophys J, 1994. **66**(1): p. 120-32.

15. Parasassi, T., et al., *LAURDAN and PRODAN as polarity-sensitive fluorescent membrane probes.* J. Fluoresc. , 1998. **8**: p. 365-373.
16. Yu, W., et al., *Fluorescence generalized polarization of cell membranes: a two-photon scanning microscopy approach.* Biophys J, 1996. **70**(2): p. 626-36.
17. Owen, D.M., et al., *Quantitative imaging of membrane lipid order in cells and organisms.* Nat Protoc, 2011. **7**(1): p. 24-35.
18. Golfetto, O., E. Hinde, and E. Gratton, *The Laurdan spectral phasor method to explore membrane micro-heterogeneity and lipid domains in live cells.* Methods Mol Biol, 2015. **1232**: p. 273-90.
19. Sanchez, S.A., et al., *Methyl-beta-cyclodextrins preferentially remove cholesterol from the liquid disordered phase in giant unilamellar vesicles.* J Membr Biol, 2011. **241**(1): p. 1-10.
20. Kubiak, J., et al., *Lipid lateral organization on giant unilamellar vesicles containing lipopolysaccharides.* Biophys J, 2011. **100**(4): p. 978-86.
21. Leung, S.S.W., et al., *Measuring molecular order for lipid membrane phase studies: Linear relationship between Laurdan generalized polarization and deuterium NMR order parameter.* Biochim Biophys Acta Biomembr, 2019. **1861**(12): p. 183053.
22. Amaro, M., et al., *Time-resolved fluorescence in lipid bilayers: selected applications and advantages over steady state.* Biophys J, 2014. **107**(12): p. 2751-2760.
23. Aguilar, L.F., et al., *Differential dynamic and structural behavior of lipid-cholesterol domains in model membranes.* PLoS One, 2012. **7**(6): p. e40254.
24. Bagatolli, L.A., et al., *Laurdan properties in glycosphingolipid-phospholipid mixtures: a comparative fluorescence and calorimetric study.* Biochimica et biophysica acta, 1997. **1325**(1): p. 80-90.
25. Bagatolli, L.A., E. Gratton, and G.D. Fidelio, *Water dynamics in glycosphingolipid aggregates studied by LAURDAN fluorescence.* Biophys J, 1998. **75**(1): p. 331-41.
26. Parasassi, T., et al., *Abrupt modifications of phospholipid bilayer properties at critical cholesterol concentrations.* Biophys J, 1995. **68**(5): p. 1895-902.
27. Mukherjee, S. and A. Chattopadhyay, *Monitoring the organization and dynamics of bovine hippocampal membranes utilizing Laurdan generalized polarization.* Biochimica et Biophysica Acta, 2005. **1714**(1): p. 43-55.
28. Schneckenburger, H., et al., *Laser-assisted fluorescence microscopy for measuring cell membrane dynamics.* Photochem Photobiol Sci, 2004. **3**(8): p. 817-22.
29. Golfetto, O., E. Hinde, and E. Gratton, *Laurdan fluorescence lifetime discriminates cholesterol content from changes in fluidity in living cell membranes.* Biophys J, 2013. **104**(6): p. 1238-47.
30. Bonaventura, G., et al., *Laurdan monitors different lipids content in eukaryotic membrane during embryonic neural development.* Cell Biochem Biophys, 2014. **70**(2): p. 785-94.
31. Malacrida, L., D.M. Jameson, and E. Gratton, *A multidimensional phasor approach reveals LAURDAN photophysics in NIH-3T3 cell membranes.* Sci Rep, 2017. **7**(1): p. 9215.
32. Gasecka, A., et al., *Quantitative imaging of molecular order in lipid membranes using two-photon fluorescence polarimetry.* Biophys J, 2009. **97**(10): p. 2854-62.
33. Parasassi, T., et al., *Two-photon fluorescence microscopy of laurdan generalized polarization domains in model and natural membranes.* Biophys J, 1997. **72**(6): p. 2413-29.
34. Wheeler, G. and K.M. Tyler, *Widefield microscopy for live imaging of lipid domains and membrane dynamics.* Biochimica et biophysica acta, 2011. **1808**(3): p. 634-41.
35. Bagatolli, L.A., et al., *Giant vesicles, Laurdan, and two-photon fluorescence microscopy: evidence of lipid lateral separation in bilayers.* Methods Enzymol, 2003. **360**: p. 481-500.
36. Bagatolli, L.A. and E. Gratton, *Two-photon fluorescence microscopy observation of shape changes at the phase transition in phospholipid giant unilamellar vesicles.* Biophys J, 1999. **77**(4): p. 2090-101.
37. Parisio, G., et al., *Polarity-sensitive fluorescent probes in lipid bilayers: bridging spectroscopic behavior and microenvironment properties.* J Phys Chem B, 2011. **115**(33): p. 9980-9.
38. Osella, S. and S. Knippenberg, *Laurdan as a Molecular Rotor in Biological Environments.* ACS Appl Bio Mater, 2019. **2**(12): p. 5769-5778.
39. Osella, S., et al., *Investigation into Biological Environments through (Non)linear Optics: A Multiscale Study of Laurdan Derivatives.* J Chem Theory Comput, 2016. **12**(12): p. 6169-6181.
40. Osella, S., et al., *Conformational Changes as Driving Force for Phase Recognition: The Case of Laurdan.* Langmuir, 2019. **35**(35): p. 11471-11481.
41. Barucha-Kraszewska, J., S. Kraszewski, and C. Ramseyer, *Will C-Laurdan dethrone Laurdan in fluorescent solvent relaxation techniques for lipid membrane studies?* Langmuir, 2013. **29**(4): p. 1174-82.
42. Wasif Baig, M., et al., *Orientation of Laurdan in Phospholipid Bilayers Influences Its Fluorescence: Quantum Mechanics and Classical Molecular Dynamics Study.* Molecules, 2018. **23**(7).
43. Suhaj, A., et al., *Laurdan and Di-4-ANEPPDHQ Influence the Properties of Lipid Membranes: A Classical Molecular Dynamics and Fluorescence Study.* J Phys Chem B, 2020. **124**(50): p. 11419-11430.
44. Kahya, N., et al., *Probing lipid mobility of raft-exhibiting model membranes by fluorescence correlation spectroscopy.* J Biol Chem, 2003. **278**(30): p. 28109-15.
45. Bianchetti, G., et al., *Investigation of the Membrane Fluidity Regulation of Fatty Acid Intracellular Distribution by Fluorescence Lifetime Imaging of Novel Polarity Sensitive Fluorescent Derivatives.* Int J Mol Sci, 2021. **22**(6).

46. Lakowicz, J.R., *Principles of Fluorescence Spectroscopy*. 3rd ed. ed. 2006, Springer: Berlin, Germany.
47. Gaus, K., T. Zech, and T. Harder, *Visualizing membrane microdomains by Laurdan 2-photon microscopy.* Mol Membr Biol, 2006. **23**(1): p. 41-8.
48. Farber, N. and C. Westerhausen, *Broad lipid phase transitions in mammalian cell membranes measured by Laurdan fluorescence spectroscopy.* Biochim Biophys Acta Biomembr, 2022. **1864**(1): p. 183794.
49. Bacalum, M., et al., *A Blue-Light-Emitting BODIPY Probe for Lipid Membranes.* Langmuir, 2016. **32**(14): p. 3495-505.
50. Abraham, M.J., et al., *GROMACS: High performance molecular simulations through multi-level parallelism from laptops to supercomputers.* SoftwareX, 2015. **1-2**: p. 19-25.
51. Braun, A.R., J.N. Sachs, and J.F. Nagle, *Comparing simulations of lipid bilayers to scattering data: the GROMOS 43A1-S3 force field.* J Phys Chem B, 2013. **117**(17): p. 5065-72.
52. M. J. Frisch, G.W.T., H. B. Schlegel, G. E. Scuseria, M. A. Robb, J. R. Cheeseman, G. Scalmani, V. Barone, B. Mennucci, G. A. Petersson, H. Nakatsuji, M. Caricato, X. Li, H. P. Hratchian, A. F. Izmaylov, J. Bloino, G. Zheng, J. L. Sonnenberg, M. Hada, M. Ehara, K. Toyota, R. Fukuda, J. Hasegawa, M. Ishida, T. Nakajima, Y. Honda, O. Kitao, H. Nakai, T. Vreven, J. A. Montgomery, Jr., J. E. Peralta, F. Ogliaro, M. Bearpark, J. J. Heyd, E. Brothers, K. N. Kudin, V. N. Staroverov, R. Kobayashi, J. Normand, K. Raghavachari, A. Rendell, J. C. Burant, S. S. Iyengar, J. Tomasi, M. Cossi, N. Rega, J. M. Millam, M. Klene, J. E. Knox, J. B. Cross, V. Bakken, C. Adamo, J. Jaramillo, R. Gomperts, R. E. Stratmann, O. Yazyev, A. J. Austin, R. Cammi, C. Pomelli, J. W. Ochterski, R. L. Martin, K. Morokuma, V. G. Zakrzewski, G. A. Voth, P. Salvador, J. J. Dannenberg, S. Dapprich, A. D. Daniels, O. Farkas, J. B. Foresman, J. V. Ortiz, J. Cioslowski, and D. J. Fox, *Gaussian 09, Revision D.01,* . 2016, Wallingford CT: Gaussian, Inc.
53. T Yanai, DP Tew, and N. Handy, *A new hybrid exchange–correlation functional using the Coulomb-attenuating method (CAM-B3LYP).* Chemical Physics Letters, 2004. **393**(1-3): p. 51-57.
54. Berger, O., O. Edholm, and F. Jahnig, *Molecular dynamics simulations of a fluid bilayer of dipalmitoylphosphatidylcholine at full hydration, constant pressure, and constant temperature.* Biophys J, 1997. **72**(5): p. 2002-13.
55. T.;, D., D. York, and L. Pedersen, *Particle mesh Ewald: An N·log(N) method for Ewald sums in large systems* J. Chem. Phys. , 1993. **98**: p. 10089–10092.
56. Nosé, S., *A unified formulation of the constant temperature molecular dynamics methods* J. Chem. Phys. , 1984. **81**: p. 511-519.
57. Hoover, W.G., *Canonical dynamics: Equilibrium phase-space distributions.* Phys Rev A Gen Phys, 1985. **31**(3): p. 1695-1697.
58. Parrinello, M. and A. Rahman, *Polymorphic Transitions in Single Crystals: A New Molecular Dynamics Method.* Journal of Applied Physics, 1981. **52**: p. 7182-7190.
59. Harris, F.M., K.B. Best, and J.D. Bell, *Use of laurdan fluorescence intensity and polarization to distinguish between changes in membrane fluidity and phospholipid order.* Biochimica et Biophysica Acta, 2002. **1565**(1): p. 123-8.
60. Benedini, L., et al., *Study of the influence of ascorbyl palmitate and amiodarone in the stability of unilamellar liposomes.* Mol Membr Biol, 2014. **31**(2-3): p. 85-94.
61. Koynova, R. and M. Caffrey, *Phases and phase transitions of the phosphatidylcholines.* Biochimica et Biophysica Acta, 1998. **1376**(1): p. 91-145.
62. De Santis, A., et al., *Omega-3 polyunsaturated fatty acids do not fluidify bilayers in the liquid-crystalline state.* Sci Rep, 2018. **8**(1): p. 16240.
63. Antollini, S.S. and M.I. Aveldano, *Thermal behavior of liposomes containing PCs with long and very long chain PUFAs isolated from retinal rod outer segment membranes.* J Lipid Res, 2002. **43**(9): p. 1440-9.
64. Kriegler, S., et al., *Structural responses of model biomembranes to Mars-relevant salts.* Phys Chem Chem Phys, 2021. **23**(26): p. 14212-14223.
65. Watanabe, N., et al., *Lipid-Surrounding Water Molecules Probed by Time-Resolved Emission Spectra of Laurdan.* Langmuir, 2019. **35**(20): p. 6762-6770.
66. Vequi-Suplicy, C.C., M.T. Lamy, and C.A. Marquezin, *The new fluorescent membrane probe Ahba: a comparative study with the largely used Laurdan.* J Fluoresc, 2013. **23**(3): p. 479-86.
67. Ameloot, M., et al., *Effect of orientational order on the decay of the fluorescence anisotropy in membrane suspensions. Experimental verification on unilamellar vesicles and lipid/alpha-lactalbumin complexes.* Biophys J, 1984. **46**(4): p. 525-39.
68. Paloncyova, M., M. Ameloot, and S. Knippenberg, *Orientational distribution of DPH in lipid membranes: a comparison of molecular dynamics calculations and experimental time-resolved anisotropy experiments.* Phys Chem Chem Phys, 2019. **21**(14): p. 7594-7604.
69. Kinosita, K., Jr., S. Kawato, and A. Ikegami, *A theory of fluorescence polarization decay in membranes.* Biophys J, 1977. **20**(3): p. 289-305.
70. Bagatolli, L., et al., *Two-photon fluorescence microscopy studies of bipolar tetraether giant liposomes from thermoacidophilic archaebacteria Sulfolobus acidocaldarius.* Biophys J, 2000. **79**(1): p. 416-25.
71. Slenders, E., et al., *Dynamics of the phospholipid shell of microbubbles: a fluorescence photoselection and spectral phasor approach.* Chem Commun (Camb), 2018. **54**(38): p. 4854-4857.
72. Haluska, C.K., et al., *Combining fluorescence lifetime and polarization microscopy to discriminate phase separated domains in giant unilamellar vesicles.* Biophys J, 2008. **95**(12): p. 5737-47.